# Evaluation of team dynamic in Norwegian projects for IT students

Salah Uddin Ahmed[a], Ingrid Sundbø[a], Jon Kvisli[a], Jon Atle Gulla[b], Letizia Jaccheri[b], Anh Nguyen-Duc[a]

[a] University of South-eastern Norway, [b] Norwegian University of Science and Technology

**Abstract**

The need for teaching realistic software development in project courses has increased in a global scale. It has always been challenges in cooperating fast-changing software technologies, development methodologies and teamwork. Moreover, such project courses need to be designed in the connection to existing theoretical courses. We performed a large-scale research on student performance in Software Engineering projects in Norwegian universities. This paper investigates four aspects of team dynamics, which are team reflection, leadership, decision making and task assignment in order to improve student learning. Data was collected from student projects in 4 years at two universities. We found that some leader's characteristics are perceived differently for female and male leaders, including the perception of leaders as skillful workers or visionaries. Leadership is still a challenging aspect to teach, and assigned leadership is probably not the best way to learn. Students is are performing well in task review, however, needs support while performing task assignment. The result also suggests that task management to be done in more fine-grained levels. It is also important to maintain an open and active discussion to facilitate effective group decision makings

## 1. Introduction

Even today there is little consensus about what a software engineering project course should cover, most academic programs have at least one such course. Some project courses cover detailed aspects, e.g. programming, usability and security issues, analysis, architecture, design or work products [1]. The main goal of such courses is to have students understand the complexity of software development, have some real-life experience, and learn to work in a team [2]. The projects should be "real" in every aspect as they entail the development of an application desired by a real world customer. As in industry, applications are developed by a small, collaborative team which needs to communicate with the customer, coordinate its activity, attend to internal decision-making, and be sensitive to delivering value [4].

Project courses complements theoretical lectures in the way that students can integrate and consolidated theory and skills through the use on project work [3]. To achieve these in academic context, students are expected to learn both technology-related knowledge and soft skills. For students, the soft skills acquired through a capstone course are perhaps one of the greatest rewards of this academic experience. These include problem solving, communication, and teamwork skills which are becoming essential for working in the industry [5]. Team dynamic is a critical component of working in small groups, yet it often goes unnoticed in the context of software engineering education. Team dynamic has been long recognized as an important success factor in teamwork. Team dynamics describes the effects of these roles and behaviors on other group members, and on the group as a whole [6]. There has been extensive body knowledge on different aspects of group dynamics, such as task interaction, relationship interaction, decision making, conflict management and group performance [7].

This research reports our experience with student team dynamics in capstone project courses at Norwegian University of Science and Technology (NTNU) and University College of Southeast Norway (USN). In both universities, we designed semester-long courses where students work with actual customers from industry. In the end of the courses, they need to deliver a prototype/ product for the customers, present the result and submit a project report. Learning to work in team is a high-priority learning objective in these courses. From 2015, we started to collect data from students with the aim at improving the courses. The data were collected from course responsible, student assistants and students in various instruments, such as questionnaires, interviews, reflection reports and workshops. In this research, we investigated students' perception on team dynamics and their concerns with four aspects, teamwork, decision-making, leadership, and task assignment. Our research questions are:

*RQ1: How do students perceive team dynamics in the end of their capstone projects?*
*RQ2: What are challenges regarding to team dynamics in such contexts?*

There is a large body of knowledge on issues and challenges pertaining to team project courses in computing education [3,8,9,10,11] and problem- and project-based learning in general [12,13]. With the focus on team dynamics, the study contributes to the current understanding about capstone projects in Software Engineering and particularly in the context of Norwegian education. We combined both quantitative data (for answering RQ1) and qualitative data (for answering RQ2). This is one of the first studies in Norway that adopts a mixed research method to study team dynamics in Software Engineering projects. The quantitative results from two universities can be used by other Norwegian universities that share the same context, to benchmark about student activity. The results also have direct implications for course design, by suggesting a better organizing and lecturing task assignment, leadership and decision making practices. The paper is organized as below. Section 2 presents related work. In section 3 we present our research methodology. Section 4 describes the settings of the course as it was conducted in the two universities. In section 5 we present the results. Section 5.1 presents the result of the survey while 5.2 presents the qualitative result addressing RQ2. Section 6 discusses and concludes the article.

## 2. Related work

We present several research about team dynamics aspects, such as team reflection, task management, decision making and team leadership as a foundation for our research. **Team reflection** is defined as "*the extent to which team members collectively reflect upon the team's objectives, strategies and processes*" [17]. Team reflection is achieved when the team members ask questions, review solutions, challenge a solution, criticize a decision, be open for alternatives and suggestions. Reflection is often done via retrospective meetings in Scrum teams. In our courses, we highlight the role of retrospective meetings as a part of course evaluation.

On student projects, **task management** involves two of the most common problems. Started tasks remained uncompleted, and tasks were planned in a vague manner [5]. Task assignment is an important task in the team. Igaki et. el describe the inequality of task assignment in a scrum student team as a challenge that makes the project assessment difficult [19]. Team effectiveness depends fundamentally upon how well team members can coordinate their actions. The factors that complicate task effort estimation can also complicate planning concrete tasks, and if tasks are not concrete, they are also more difficult to complete. Tasks remaining open may also indicate the lack of regular effort invested in the project. We have often seen that some students or even some teams increase their weekly effort to the project only closer to the end of the phases or the end of the project, if ever.

Most of the research conducted prior to 1990s distinguished between two approaches to **leadership**: task-oriented style, defined as a concern with accomplishing assigned tasks by organizing task-relevant activities, and interpersonally oriented style, defined as a concern with maintaining interpersonal relationships by tending to others' morale and welfare [20]. Compared to industrial placement the learning in a capstone project is higher, as the students can get involved in more phases and have greater roles than in an industrial placement. Even though it might seem to be a bit hard for a leader of the capstone project to get his/her team motivated when the expectations and effort varies widely in a student team, the leadership position is still rewarding. Vanhanen, et al. report that students working in the managerial roles on the average, learned quite a lot about software engineering topics compared to the developers who learned moderately [5].

Participative **leadership** was positively associated with the process of team reflection for highly functional heterogeneous teams [18]. Heterogeneous team is a team where the members have different levels of skills, knowledge and abilities. However in case of a homogeneous team where the members have similar levels of background knowledge and skills, directive leadership instead of participative is desired for invoking team reflection. One important role for the participative leader in heterogeneous teams is to help team members translate the advantages of heterogeneity, such as the variety of professional backgrounds, knowledge, skills, and abilities, into significant processes of questioning, reviewing, and exploring [17]. In homogenous teams, because of the similarity of the members questioning, reviewing and exploring does not happen naturally, therefore, a directive leader can introduce them by enforcing.

## 3. Research methodology

Research design was conducted in 2015 for research-informed teaching approaches in customer driven project courses. We collected various data about (1) leadership, (2) female participation in projects (3) team dynamics. The study occurs firstly in the context of course TDT4290 Customer Driven Project at NTNU. The course has been taught since 2011, in which students in their fourth year need to develop a software product/ service for a real customer. The second course is PRO1000B, which is taught for $1^{st}$ year students at University of Southeast Norway (USN). Students also work for a real customer in making a prototype demonstrating their business ideas.

The data collection periods was done from Aug 2015 to Mar 2018. Various instruments were used to collect data about how teams perform during their projects, including project plans, final reports, supervision meeting notes, interviews with team leaders, and team reflection survey. In this study, we used two main instruments that reflect team dynamics, which are team reflection survey and final reports.

1) We designed the team reflection survey to collect team's perception on their own team dynamic. The survey used a five-point Likert scale to collect leaders and team member's opinions on (1) their own performance, (2) collective decision making, (3) team leadership and (4) task management practices.

2) Final project report. Each team delivered a 150-200 pages project report describing project planning and management, product requirement and architecture, testing and delivery. Especially, a final part of the report consisted of team reflection on project mandate, teamwork and supervision.

Table 1: Project course setting

| Course | Year | No. Groups | Student background |
|---|---|---|---|
| **TDT4290** | 2015 | 12 | $4^{th}$ year students from Information Technology, Computer Science, Erasmus students, … |
| **TDT4290** | 2016 | 13 | |
| **TDT4290** | 2017 | 14 | |
| **PRO1000B** | 2018 | 11 | $1^{st}$ year students from IT, Marketing, etc |

We provide a statistic descriptive of feedbacks from 36 students groups ( total 163 students, N=163). The survey includes 18 questions with a 5-point Likert scale (strongly agree - strongly disagree). Each question has a N/A (Not applicable) option if students do not experience the asked situations. For each group we take median values of all group members to represent for the group dynamics. Then we describes the answers of group feedbacks using boxplots. We also performed two-tail t-test to see if there is a different perception among students who belongs to categories of different universities, study year, and leader's genders.

While the survey provides an overall assessment on team performance, they do not provide details on area of improvements for students. We complemented for this gap by investigating student reflection reports. We performed a tailored thematic analysis [14], a qualitative approach for analyzing data from interview transcripts, observation notes and documents. We extracted text from student reports about team dynamics, especially about team decision making, leadership and task management. The texts are labelled and put into either of three investigated category. It is noted that not all of the groups report on these aspects of team dynamics, hence we do not do a cross-team comparison. In the end, a list of possible challenges were created as a thematic map.

## 4. Course setting

The course was designed in a similar manner, including five elements (1) Customer, Team assignment, (3) Grading, (4) Supervision and (5) Established Software development paradigm. In both of the courses, we invited developers, managers, entrepreneurs from companies and research institutes to be **customers**. They often have some ideas or ongoing project that needs some research and development tasks. During the projects, each customer actively participates in the requirements definition work as well as monitors the project progress. The customer must prune the scope of the project to fit to the number of working hours students can spend on the course.

Students are **assigned** in project groups of 6-8 students. At USN, we adopted self-selected team approach, which allow students to have more control over project development. At NTNU, we adopted both random assignment and instructed selection. Random assignment makes groups where the members generally do not know each other beforehand. The project work will be **graded** on the basis of the quality of the project report, the functioning prototype of the system, and the presentation delivered at the end of the course. The team dynamics will have an impact on the final grade.

Each team is assigned a **supervisor**, who will assist the team throughout the courses. The supervisor can help with pedagogical concerns, facilities for project execution, concerns with team and customers and probably technical issues. The courses maintain a regular supervision meeting to discuss current matters for the teams and to update status. Supervisors also help with giving evaluation on team performance in the end of the courses.

Students are introduced to fundamental **software development life cycles**, namely Waterfall, Iterative development, Agile and Lean Startup. In general, prior to the course, students have learnt about software engineering processes and practices. At NTNU, students are free to choose the methods that fit to their projects. In some cases, the development approaches are enforced by customers, due to the existing working process they adopt in their organizations. At HSN, the students are guided to apply Scrum in all groups. A series of lectures about Scrum development was given in the early phase of the course.

# 5. Result

## 5.1. RQ1: How do students perceive team dynamics in the end of their projects?

We provide a summary of mean values for each questions rated by students as in Figure 1. We found that the top three items that gain the highest student rates are Q1, Q4 and Q18. Students believe that they propose more than one alternatives for a problem to discuss, which reflects an open environment for discussion. Students also feel they are free to criticize other opinions. Among task management practices, task review is perceived to be best performed. We found the top three items that have lowest rates from students, which are Q7, Q15, and Q16. Teams feel that they are not confident with problem solving skills. Moreover, they have challenges with assigning time and responsibilities for tasks.

Figure 1 also reveals the elements that have largest variation in students' opinions. These all belong to the reflection on leadership: Q10 (perception of a leader as a visionary), Q12 (perception of a leader as a skillful worker) and Q14 (perception of a leader as an exciting public speaker). The most stable elements are student's ability to look for different interpretations and to evaluate their weaknesses.

Each team is characterized by their study location (USN or NTNU), their leader gender (male vs. female) and their course year (2015, 2016 and 2018). We adopted two tailed t-tests for all survey questions (from Q1 to Q18 as shown in Table 3) to compare the features of students in different categories. The t-test with significant p-value at 95% is described in **Error! Reference source not found.**.

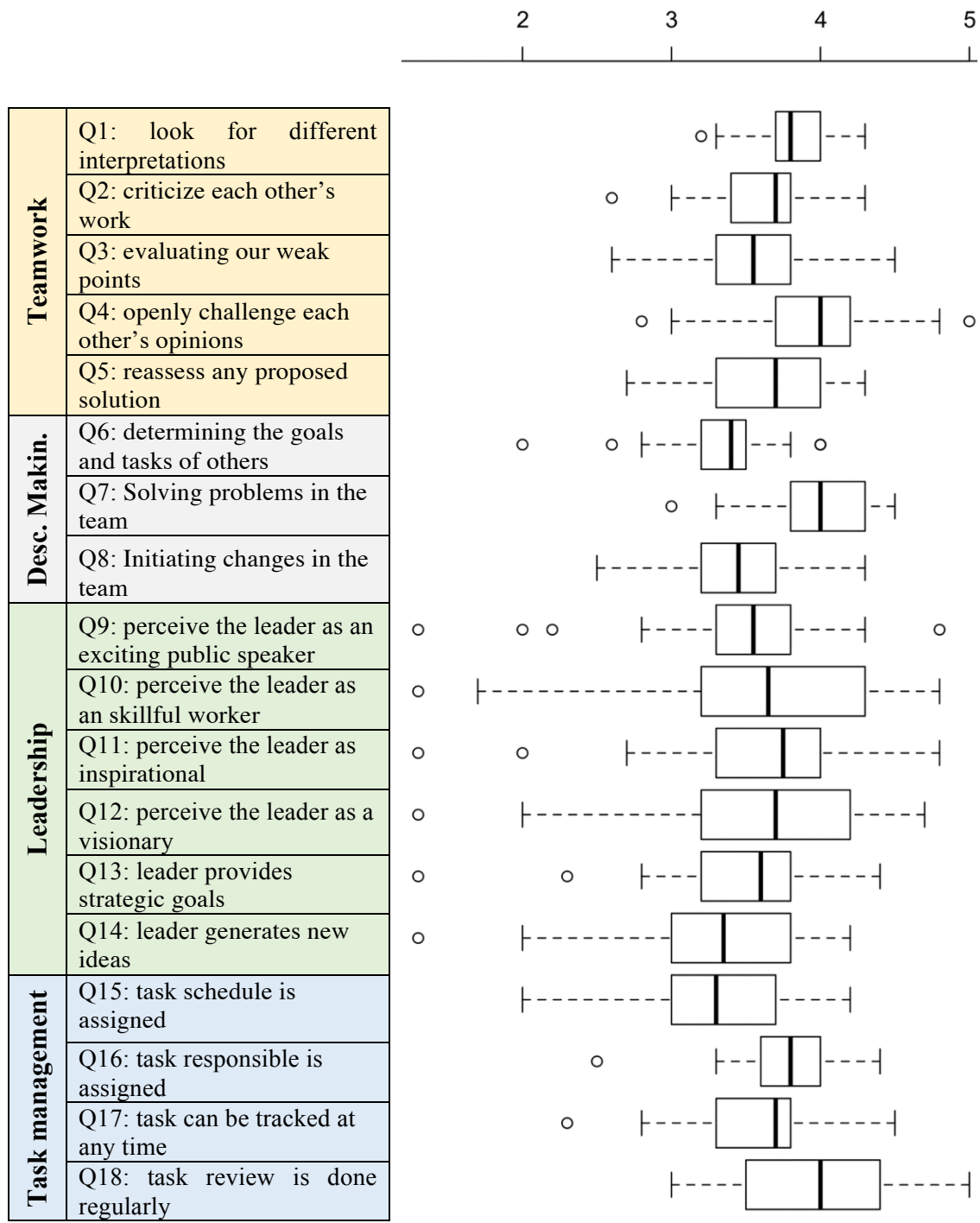

Figure 1: Boxplots of student rates on team dynamics

According to t-test results, there is a significant difference on how student evaluate their weakness between USN and NTNU courses, with the mean value of USN students (3.24) slightly smaller than of NTNU students (3.57). Two aspects of team decision making have been perceived differently between students in 2015 and students in 2016. These are decision of goals and task for other members (3.46 vs. 3.05) and solving team problems (4.1 vs 3.8). Interestingly, both of the student rates for 2016 is lower than the ones for 2015.

Table 2: Comparison across course year, campus and leader gender

| Hypotheses | Survey Ques. | P value |
|---|---|---|
| Students in USN and NTNU perceived their team dynamics differently | Q3 | 0.0256 |
| Students perceived their team dynamic differently among 2015, vs 2016 at NTNU | Q6 | 0.0171 |
| | Q7 | 0.0461 |
| Students perceived their team dynamic differently among female-leading team and male leading team | Q3 | 0.0136 |
| | Q10 | 0.0500 |
| | Q12 | 0.0487 |
| | Q18 | 0.0434 |

Regards to leader's gender, we identified four aspects that student perception might be different between female leader teams vs male leader teams. Two questions are related to how leader perform as a skillful worker (Q10) and as a visionary (Q12).

## 5.2. RQ2: What are challenges regarding to capstone team dynamics in such contexts?

Figure 2 describes the list of concerns regarding students' reflection on team dynamics. We present different concepts found in sections below.

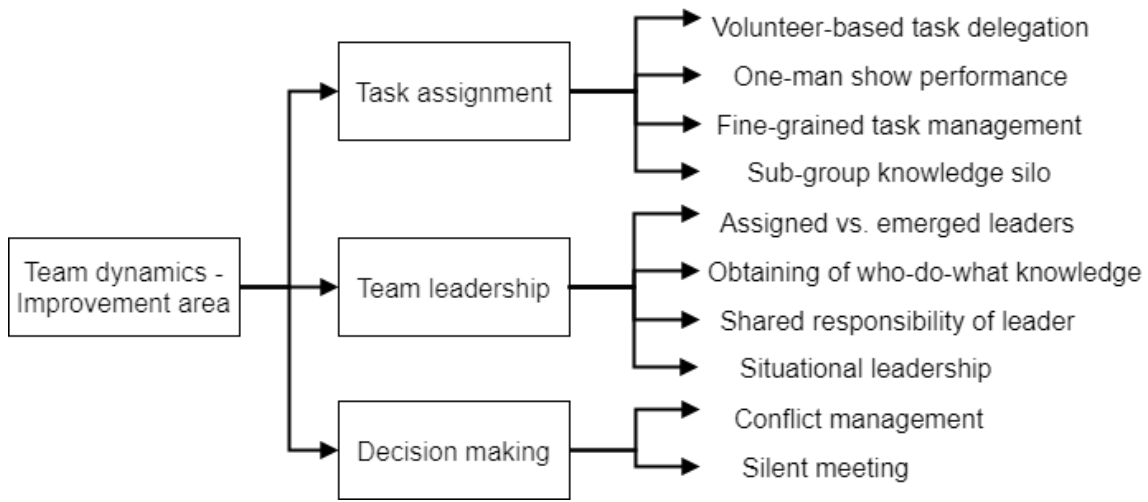

Figure 2: Thematic map of team dynamics concern

*5.2.1. Task assignment:*

**Volunteer-based task delegation.** Students understand the importance of planning and delegating work for ensuring an internal workflow. It is common that teams assign themselves to tasks based on volunteering. As mentioned by a team: "*Among the many things that went well in this project was the division of tasks between members. As mentioned, everyone had different experiences and interests, and the team tried to delegate tasks fairly so that everyone got to work on something that they were interested in. The team also took the time to make sure that everyone was happy with the tasks they were assigned at the beginning of a sprint.*" (NTNU15G04). Teams evolve into a set of functional roles, as stated by a group: "*Members were encouraged to focus on tasks they were good at, to increase efficiency and productivity. This led to some members working a lot more on the report than on development.*" (NTNU17G01). The concern here is that some students will never learn other skills. For instance, some students may feel comfortable with documentation, writing emails, and contacting customers, hence they will have less chance to learn technical knowledge. Other students, who may focus on coding, will less likely be involved in customer meeting, and market research.

**One-man show**. The balance of teamwork can be effected by a "star" in the team. A student groups mentioned: "*We were afraid that the people that were experienced, with the technologies we decided upon, would start setting up the system and implementing features on their own. That could lead to other members being demotivated. One of our team members knew a lot about our technologies. However, after slightly mentioning the concern discussed here, our knowledgeable member was more than willing to spend time to help us get to know the technologies ourselves*" (NTNU15G05). We observed that there is always a role of "technical expert" in a project team. There is a risk associated with the knowledge gap between team members: "*The different skill levels with the chosen technology going into the project led to a bit of pair programming and tutorial-like workshops, especially in the beginning of the project. At times it could take too much time from someone got stuck on a task until it was reassigned or finished with help*" (NTNU17G03). Managing communication between this member and the rest of the team is important for moving from a toy exercise to an actual collaborative work.

**Fine-grained task management.** Team members might not deliver works on-time due to many reasons, such as sickness, deadlines from other courses, travel, etc. This could be a critical issue for inter-dependent tasks. Tasks like coding might need an approach that is more frequent in synchronizing tasks. Taking a look at the quote from a team report: "*During the project we noticed that instead of having a general deadline at the end of the sprint, having deadlines for tasks during the sprints made a big difference in our productivity. By using specific deadlines tasks got done evenly through the sprint and not at the last couple of days, which meant that we had more time to test the whole product before sprint reviews. By distributing tasks throughout the sprints we also spread out the pull requests which resulted in less merge conflicts.*" (NTNU17G09). We recalled many circumstances that students need to wait for several weeks to get into the next steps of their projects due to a deadlock wait for a task to be completed. While the reasons might be external dependencies, hardware and device dependencies, in general there should be a better approach in dividing and managing such tasks.

**Sub-group silo**. There is also a common practice to maintain functional sub-groups "*We believe it was necessary to divide into sub-teams in order to complete the different parts of the project. Having everyone involved in all the different parts would incur too big of an overhead, because many of the technologies took significant time getting into*" (NTNU17G09). This is observed in several groups "*Even though some team members have preferred to work individually, the collaboration within the group divisions have worked well.*" (NTNU15G01). More frequent meetings can be hosted in sub-teams and require less meetings for the whole team. Without a proper adoption of project management tool, we found that a team member in a sub-group might not be aware of who-do-what knowledge in other sub-groups.

*5.2.2.Leadership:*

**Assigned vs. emerged leaders.** At NTNU, for the courses in 2016 and 2017, we assigned a leader for each team. Although this could reflect a real-world situation where students could be put into an on-going project, this has generated some challenges in the course context. In many cases, the assigned leaders do not express a clear responsibility: "*The*

*fact that we could not control who became the leader of the project was a bit strange. In a group dynamic this can be very damaging for collaboration. If the leader is not invested enough in the project, or very uncomfortable it might damage the process aspect of the project.*" (NTNU17G06). "*... The team believes that it might be better to allow the groups to decide how to choose their leader themselves next year, although this method has its drawbacks as well.*" (NTNU17G09). In team NTNU17G07, it appeared a role of an emerged leader, who influenced the team regarding to decision-making, task assignment and collective problem solving. In this case, the emerged leader participated in leadership meetings instead of the assigned leader. The issues of assigned vs. emerged leadership do occur in case of autonomous teams as well. Team USN18G05 initially agreed to have one team member to be a leader. The initial leader with an authoritarian style establishes team contracts, meeting schedule and contact to customers. Later, the team is more influenced by a technical leader, who also acts as a Scrum Master. The team adopts Scrum approach and less rely on mechanistic task assignment

**Obtaining of who-do-what knowledge**. Many students have their first time experiences of being leaders. Most of the leaders in the first Sprint have challenges of understanding task flow and team responsibility: "*In the beginning the leader did not follow up enough; asking members what they were working on, how it was going, if they needed help or new tasks. This reasoning is based on different perceptions between us, where the leader expected the members to take initiative to provide a status update, whereas the members expected the leader to follow up*" (NTNU17G09). Some teams do not utilize project management tool, such as Trello, Jira, effectively for visualizing project progress.

**Shared responsibility of leader.** A team reported a risk of unfulfilled leader's responsibility leading to teamwork delay: "*As a practical note, the team should have appointed a vice-leader in case the team leader became indisposed. This turned out to be an issue when the team leader became ill with pneumonia. Her absence meant that the team's communication with the customer and communication within the team took a hit during that period*" (NTNU17G07). In team USN18G04, the team leader quit the project after five weeks, leading to a significant impact on team progress. The leader was responsible for contacting customers and maintaining the collaborative workspaces. All of this information was not transferred to other team members.

**Situational leadership.** According to the theory of leadership, the leadership style, which is fit to a task context, will provide the best team performance [15]. Students work on different projects with different level of requirement clarity, task dependencies, and portion of research [16]. Besides, different teams have their own characteristics, such as the familiarity among team members, control and coordination mechanisms [16]. Even though we have not gathered sufficient data about task context at this stage, we can hypothesize that the leadership style should be situational in reflecting the team and task situation. Hence, assigning a team leader with a task-oriented styles to a team that lacks of task structure might be a challenge for the team.

*5.2.3.Decision making:*
**Conflict management.** Decision making practices are directly influenced by leadership styles. Individuals would either identify a team leader, who has a final word, or else claim that their team has no leader and every member has an equal role in guiding team decisions: "*If a decision was to be made by a vote, the group leader had a double-vote during ties. This was to ensure that a decision was made even if there were to be a tie, as*

*there was an even number of group members"*. (NTNU15G12). When asked how their team made decisions, many participants described a very democratic style of decision making in which each member had a voice, all opinions were respected, and the group made decisions together. A threat for democratic team is conflicts among team members. The most problematic conflict is probably member commitment. Conflict of priorities between the team member's personal tasks and group work makes it difficult for members to commit and deliver the promised work. Teams that successfully manage conflicts would perform a good team dynamic: *"If there was ever a disagreement or any specific cases to discuss, this was always brought up in the meetings. Usually critiques or other feedback about the group dynamics were given in retrospective meetings. This strengthened the teamwork and gave each member something to improve on for both this specific project, but also an experience for later projects"* (NTNU17G02)

**"Silent meeting"**. It is important to establish an open environment where team members can throw their opinions on the table: *"Different ideas and thoughts were well received by everyone on the team, which helped to create open discussions and a safe environment to share opinions."* (NTNU17G09). One way to trigger discussion and idea generation is to assign roles for each team members: *"At the start of the project the group decided to assign roles to the various team members in order to ensure that the project would proceed as planned. This was a huge benefit to the project as assigning roles to the team members gave them an area of responsibility. Being liable for a part of the project gave the responsible individual an incentive for ensuring that the necessary work was done, and that tasks were delegated as needed"*

## 6. Discussion and Conclusions

Team work is one of the prime software engineering concept that a student learn from a capstone course. Good team dynamics is directly related to project success and customer satisfaction. In order to teach the students team dynamics, it is important to find out what constitutes team dynamics. Team dynamics can vary based on context and locale. In this paper we have tried to find out the components of team dynamics from the literature and gained an insight about the team dynamics in context of Norwegian universities from the survey and student reports. Several findings are observed:
- There is a large variation in perception on leaderships and leadership styles among students
- Some leader's characteristics are perceived differently for female and male leaders, including the perception of leaders as skillful workers or visionaries
- There might be issues with the assigned leaders
- Task review is well perceived, however, task assignment needs further care.
- Fine-grained task management rather than Sprint-based milestones
- Maintaining an open and active discussion to facilitate effective group decision making

Several of our findings matches with the values found in the literature. For example, emerged leader instead of assigned leader tends to perform better increasing the team performance. We also find that students are in general pleased with project-based courses due to their learning and simulated real world experiences [2,5]. Task assignment is still a challenge for students [19]. We found that there should be a complement mechanism to support students assigning and managing tasks besides Sprint planning. Despite of the importance of leadership [5,17,20], we found that there should be more focus on teaching students principles of situational leaderships.

The results of this study can be used to improve the project courses involving real customers. The survey result informs the areas that students are not confident with, i.e. task assignment and leadership. We suggest to presentation of our findings to prepare students aware of possible threats. We also suggest that course responsible should survey students' working styles and previous knowledge and experience in teamwork. Information about customers' projects should be analyzed and categorized based on their task clarity, difficulty and level of customer involvement. These information should be used to assist the assignment of leaders into teams, for the best team learning. Moreover, in the connection to the theory part of such courses, we suggest that development paradigm should be taught thoroughfully. Hand-ins exercises for collaborative practices, i.e task assignment, effort estimation, etc would be beneficial for many course scenarios.

There are some threats to validity that worth to discuss [21]. A threat to construct validity is the relation between theory and our observation. Our data were collected in several ways, i.e. surveys, interviews and supervisors' observation notes. The data collection instruments were based on existing research, which ensures our credibility. Team dynamics includes many aspects that we do not focus on, i.e. team communication, team culture, team cohesion and social identity. We limited the scope of this study on four aspects, team reflection, decision making, task management and team leadership. A threat to external validity concerns how we can generalize the findings. Comparing to literature, we confirms some challenges with students learning in modern software development project setting. As the study was conducted in more than one universities, we believe the result can be applicable for European universities' programs with similar settings.

This study presents an initial step for a mixed research approach on studying student software projects. Further steps would include more data from other courses to confirm our findings. The study barely touches the concept of gender participation in software projects. Our next work will report the understanding on female participation and leadership in this context. A comprehensive investigation on team dynamics' aspects will generate a guideline for students and curriculum design.

# References

1. J. Borstler. "Experience with work-product oriented software development projects". *Computer Science Education*, vol 11(2), pp. 111–133, 2001
2. R. Lingard and S. Barkataki. "Teaching teamwork in engineering and computer science". In Frontiers in Education Conference. IEEE, Rapid City, SD, F1C pp 1-5, 2011
3. S. Fincher, M. Petre, and M. Clark. *Computer Science Project Work: Principles and Pragmatics*. Springer Science & Business Media, London, UK, 2001
4. P. J. Denning and R. Dunham, "Innovation as language action", *Communication of ACM*, vol. 49(5), pp. 47-52, 2006.
5. J. Vanhanen, T. O. A. Lehtinen, and C. Lassenius, "Software engineering problems and their relationship to perceived learning and customer satisfaction on a software capstone project", *Journal of Systems and Software*, vol. 137, pp. 50-66, 2018/03/01/, 2018.
6. K. Lewin, *A dynamic theory of personality*. New York: McGraw-Hill, 1935
7. D. R. Forsyth, *Group Dynamics*, 4th Edition, Wadsworth Publishing, 2005

8. I. Crnković, I. Bosnić, and M. Zagat. "Ten tips to succeed in global software engineering education" *34th International Conference on Software Engineering.* pp. 1225–1234, 2012
9. J. C. H. Ellis, S. A. Demurjian, and J. F. Naveda. "Software Engineering: Effective Teaching and Learning Approaches and Practices" *IGI Global,* Hershey, NY, 2009
10. T. B. Hilburn and W. S. Humphrey. " Teaching Teamwork". *IEEE Software* 19, 5, pp. 72–77, 2002
11. G. Wikstrand and J. Borstler. " Success factors for team project courses" In *Proceedings of the 19th Conference on Software Engineering Education and Training.* pp. 95–102, 2006
12. L. Helle, P. Tynjala , and E. Olkinuora,"Project-based learning in post-secondary education— theory, practice and rubber sling shots", *Higher Education* 51, 2, pp. 287–314, 2006
13. C. E. Hmelo-Silver. "Problem-based learning: What and how do students learn?"*Educational Psychology Review, vol* 16(3), pp. 235–266, 2004
14. D. S. Cruzes and T. Dyba, "Recommended Steps for Thematic Synthesis in Software Engineering", 2011 International Symposium on Empirical Software Engineering and Measurement (ESEM '11), 2011
15. F. E. Fiedler, "A contingency model of leadership effectiveness", In: Berkowitz L, ed. *Advances in Experimental Social Psychology.* New York, NY: Academic Press; pp.149-190, 1964
16. A. Nguyen-Duc, S. Khodambashi, J. A. Gulla, J. Krogstie, P. Abrahamsson , "Female Leadership in Software Projects—A Preliminary Result on Leadership Style and Project Context Factors" In: Kosiuczenko P., Madeyski L. (eds) Towards a Synergistic Combination of Research and Practice in Software Engineering. Studies in Computational Intelligence, vol 733, 2018
17. M. A. West, "Reflexivity and work group effectiveness: A conceptual integration" In M. A. West (Ed.), Handbook of Work Group Psychology (pp. 555-579). Chichester: John Wiley & Sons Ltd, 1996
18. S. J. Zaccaro, A. L. Rittman, M. A. Marks, "Team Leadership", *The Leadership Quarterly* vol. 12(4), pp. 451-483, 2001
19. H. Igaki, N. Fukuyasu, S. Saiki, S. Matsumoto, and S. Kusumoto, "Quantitative assessment with using ticket driven development for teaching Scrum framework" 36[th] Inter. Conf. on Software Engineering,, Hyderabad, India, pp. 372-381, 2014
20. A. H. Eagly, and L. L. Carli, "The female leadership advantage: An evaluation of the evidence" *Leadership Quarterly*, 14, 807–834, 2003
21. R. Feldt and A. Magazinius, "Validity Threats in Empirical Software Engineering Research - An Initial Survey", 22nd International Conference on Software Engineering and Knowledge Engineering, 2010